\documentclass[12pt]{article}
\begin{document}
\title{On the derivation of the spacetime metric from linear electrodynamics}
\author{Andreas Gross\thanks{Email address: ag@thp.uni-koeln.de}\ \ and 
Guillermo F. Rubilar\thanks{Email address: gr@thp.uni-koeln.de} \\
\ \\
Institute for Theoretical Physics \\
University of Cologne\\ 
D-50923 K{\"o}ln, Germany}
\maketitle
\begin{abstract}
In the framework of metric-free electrodynamics, we start with a {\em
linear} spacetime relation between the excitation 2-form $H = ({\cal D}, 
{\cal H})$
and the field strength 2-form $F = ({E,B})$. This linear relation is 
constrained by
the so-called closure relation. We solve this system algebraically and
extend a previous analysis such as to include also singular
solutions. Using the recently derived fourth order {\em Fresnel}
equation describing the propagation of electromagnetic waves in a
general {\em linear} medium, we find that for all solutions the fourth
order surface reduces to a light cone.  Therefrom we derive the
corresponding metric up to a conformal factor.
\end{abstract}

\section{Introduction}
Recently, a metric-free formulation of the classical electromagnetic 
theory and its 
axiomatics has been discussed by Hehl and Obukhov, see
refs.\ \cite{metric1,book}. 
The basic quantities in this formalism are the electromagnetic excitation 
2-form $H$, directly related to the sources of the field, and the field 
strength 2-form $F$, describing the effects of the electromagnetic field 
on electrically charged test currents. The Lorentz force density 
$f_\alpha=(e_\alpha \rfloor F)\wedge J$ acts on test currents, were 
$J$ is the electric 
current 3-form. The $2$-forms $H$ and $F$ satisfy Maxwell's equations, 
${\rm d}H=J$, ${\rm d}F=0$. Field 
strength and excitations are not independent, but related by the so-called 
spacetime relation $H=H(F)$; see refs.\ \cite{book,foundem,post} for details.

The spacetime relation, expressing the properties of the underlying
spacetime, will in particular determine the dynamical properties of the 
electromagnetic field. In this way, it is the analogous of the constitutive
law, determining the the dynamical properties of the electromagnetic field
\emph{in material media}. The correspondence between electrodynamics in a
gravitational field and in material media have been studied before, see for 
instance refs.\ \cite{gordon, leonhardt}.

Here we consider the case of a {\em linear spacetime relation}, where $H$ is 
proportional\footnote{Recently, some nonlinear models have been discussed in a
similar formalism by Lorenci et al.\ \cite{lorenci, lorenci2}.} 
to $F$. In terms of coordinate components, the field strength
will be denoted as $F=(1/2)\,F_{ij}\, {\rm d}x^i\wedge {\rm d}x^j$ and the
excitation 2-form as $H=(1/2) \, H_{ij} \, {\rm d}x^i\wedge {\rm d}x^j$. 
Then, the most general linear spacetime 
relation can be written in terms of the {\em electromagnetic spacetime tensor 
density} 
$\chi^{ijkl}$ (of weight $+1$), as\footnote{We denote by $\epsilon_{ijkl}$ 
the completely antisymmetric tensor density of weight $-1$ with 
$\epsilon_{0123}:=1$.}
\begin{equation}\label{dual}
H_{ij}=\frac{f}{4}\,\epsilon_{ijkl}\,\chi^{klmn}\,F_{mn}, 
\qquad i,j=0,1,2,3 \, ,\label{linear}
\end{equation}
where $f$ is a dimensionfull scalar such that $\chi^{ijkl}$ is 
dimensionless.
The spacetime tensor density has the following symmetries:
\begin{equation}
\chi^{ijkl}=-\chi^{jikl}=-\chi^{ijlk}=\chi^{klij} \, .\label{symmchi}
\end{equation}

Note that the above formalism is metric-independent. As a particular 
example, however, one can recall the case of the coupling of the 
electromagnetic field to gravity in Einstein's theory, i.e.\ minimal coupling 
to the metric tensor $g_{ij}$. This corresponds to the following particular 
electromagnetic spacetime tensor density \cite{post},
\begin{equation}\label{chig}
\stackrel{\rm g}\chi{}^{ijkl}:=\sqrt{-g}\left( g^{ik}g^{jl}-
g^{jk}g^{il}\right) \, ,
\end{equation}
which we do \emph{not} assume to hold. Rather, we want to derive this
equation from the
linear ansatz (\ref{linear}).

\section{Closure relation}
With the spacetime tensor available, we may define a duality operator 
$^\#$ acting on 2-forms by extending (\ref{dual}) 
to any 2-form $\Theta=\frac{1}{2}\Theta_{ij} \, {\rm d}x^i\wedge{\rm d}x^j$ 
such that
\begin{equation}
^{\#}\Theta:=\frac{1}{4}\, \epsilon_{ijkl} \, \chi^{klnm} \, 
\Theta_{mn} \, {\rm d}x^i\wedge{\rm d}x^j \,.
\end{equation} 

In refs.\ \cite{metric1,book,metric} spacetime tensors satisfying a 
so-called {\em closure relation} have been studied, namely those with
\begin{equation}\label{cr1}
^{\#\#}=-1\, .
\end{equation}
This additional condition can be motivated by considering the 
e\-lec\-tric-mag\-ne\-tic reciprocity of the energy-momentum 
current\footnote{The energy-momentum 3-form 
of electrodynamics, $\Sigma_\alpha=(1/2)\,[ F\wedge\left(e_\alpha\rfloor
H\right)-H\wedge\left(e_\alpha\rfloor F\right)]$, 
is explicitly symmetric under a
transformation $F\rightarrow \phi H$, $H \rightarrow -(1/\phi) F$, 
for an arbitrary pseudo scalar $\phi$; see also
refs.\ \cite{book, energy-momentum, mielke} 
for a detailed discussion.}. Indeed, the closure relation results from assuming
(\ref{linear}) to be electric-magnetic reciprocal, provided we 
choose $f$ such that $f^2\phi^2=1$. Note also that in the 
particular case (\ref{chig}) this condition is only fulfilled for metrics 
with Lorentzian signature.

In order to find the solutions of (\ref{cr1}), it is convenient 
to adopt a more compact {\it bivector} notation by
defining the indices $I,J,\dots = 01, 02, 03, 23, 31, 12$.
In this notation, $\chi^{ijkl}$ corresponds to a symmetric
$6\times6$ matrix $\chi^{IJ}$ and the totally antisymmetric
$\epsilon$-tensor density becomes 
\begin{equation}
\epsilon^{IJ}=\epsilon_{IJ}=
\left(\begin{array}{rr}
0_3&1_3\\
1_3&0_3
\end{array}\right)\,.
\end{equation}
We define a new matrix $\kappa$ by 
\begin{equation}
\kappa_I{}^J:=\epsilon_{IK}\, \chi^{KJ}\,.\label{kappa}
\end{equation}
As $\epsilon^{IK}\, \kappa_K{}^J=\chi^{IJ}$, any solution of the
closure relation is given by a real $6\times 6$ matrix $\kappa$ fulfilling 
\begin{equation}\label{cr2}
\kappa_I{}^K\, \kappa_K{}^J=-\delta_I^J \label{kappa2}\,,
\end{equation}
provided $\epsilon^{IK}  \kappa_K{}^J$ is symmetric.

We decompose the matrices $\chi^{IJ}$ and $\kappa_I{}^J$ into $3\times3$ 
block-matrices,
\begin{equation}
\chi^{IJ}=\left(\begin{array}{rr}
A&C\\
C^T&B
\end{array}\right)\label{block}\,,\quad\qquad
\kappa_I{}^J=
\left(\begin{array}{rr}
C^T&B\\
A&C
\end{array}\right)\,,
\end{equation}
with symmetric matrices $A$ and $B$, and $^T$ denotes matrix transposition.

Then the closure relation, (\ref{cr1}) or (\ref{kappa2}), translates into 
\begin{eqnarray}
C^2+AB&=&-1_3\,,\label{eins}\\
BC+C^TB&=&0_3\,,\label{zwei}\\
CA+AC^T&=&0_3\,.\label{drei}
\end{eqnarray}

A consideration of the following disjoint subcases will provide the general
solution of the closure relation: 1) $B$ 
regular, 2) $B$ singular, but $A$ regular, and 3) $A$ and $B$ singular.

\section{Solutions of the closure relation}
\subsection{$B$ regular}
\noindent
We construct the general solution of (\ref{eins})-(\ref{drei}) for the 
case in which $\det B\neq 0$. Under these conditions (\ref{zwei}) is solved 
by 
\begin{equation}
C=B^{-1}K\,, \quad K^T=-K \, ,\label{c1}
\end{equation}
with an arbitrary antisymmetric matrix $K$. Using this solution for $C$, we 
rewrite (\ref{eins}) as
\begin{equation}
(B^{-1}K)^{2}+AB=-1_3 \, ,
\end{equation}
so that the solution for $A$ is given by
\begin{equation}\label{a1}
A=-B^{-1}\left[1+(KB^{-1})^{2}\right] \, .
\end{equation}
The symmetry of $A$, as assumed in (\ref{block}), may easily be seen in eq.\ 
(\ref{symma}) below.

A short calculation shows that the solutions (\ref{c1}) and (\ref{a1}) satisfy 
also (\ref{drei}) identically:
\begin{eqnarray}
CA+AC^T&=&(B^{-1}K)\left[-B^{-1}(1+(KB^{-1})^2)\right]\nonumber\\
&&+\left[-B^{-1}(1+(KB^{-1})^2)\right](-KB^{-1})\nonumber\\
&=&-B^{-1}KB^{-1}-B^{-1}KB^{-1}(KB^{-1})^2\nonumber\\
&&+B^{-1}KB^{-1}+B^{-1}(KB^{-1})^2KB^{-1}\nonumber\\
&=&0\,.
\end{eqnarray}
Therefore, the solution of the closure relation can be written as
\begin{equation}\label{s1}
\chi^{IJ}=\left(
\begin{array}{cc}
-B^{-1}\left[1+(KB^{-1})^{2}\right]&B^{-1}K\\
-KB^{-1}&B
\end{array}
\right)\,,
\end{equation}
provided $\det B \neq 0$.
This solution consists of 9 independent parameters: 3 from the antisymmetric 
matrix $K$ and 6 from the nonsingular symmetric matrix $B$.

The solution previously found in ref.\ \cite{metric}, eq.\ (14), is equivalent 
to ours, as a simple calculation using computer algebra shows 
\begin{equation}\label{symm}
A=-B^{-1}(1+(KB^{-1})^{2})=pB^{-1}+qN \, ,\label{symma}
\end{equation}
with $K=(K_{ab})=(\epsilon_{abc}k^c)$, $N=(N^{ab})=(k^ak^b)$,
$q=-1/\det{B}$,
and $p=-1+{\rm tr}(NB)/\det{B}$.

\subsection{$B$ singular, $A$ regular}

In this case (\ref{s1}) is not valid, but one can find a 
solution following a similar procedure, now starting with an arbitrary
non-singular matrix $A$. We solve (\ref{drei}) with respect to $C$ and then 
use that result to solve (\ref{eins}) with respect to $B$. In this 
way the solution can be found to be
\begin{equation}
\chi^{IJ}=\left(
\begin{array}{cc}
A & L A^{-1}\\
-A^{-1}L & -A^{-1}\left[1+(LA^{-1})^{2}\right]
\end{array}
\right)\,.\label{s2}
\end{equation}
Here $L$ is again an arbitrary antisymmetric matrix, with components 
$L=(L^{ab})$, $L^{ab}=: \epsilon^{abc} l_c$.

This solution has $8$ independent parameters: $3$ from the 
antisymmetric matrix $L$ and 6 from the nonsingular symmetric matrix $A$,
fulfilling the constraint\footnote{Note, this solution is also valid for the
case $\det{A}\neq 0$ \emph{and} $\det B \neq 0$. But in this case,
(\ref{s2}) is just a reparametrization of (\ref{s1}).} $\det{B}=0$.
The constraint reads explicitly
\begin{equation}
\det{B}=-\det{\left(A^{-1}\left[1+(LA^{-1})^{2}\right]
\right)}=0 
\,,
\end{equation}
or, equivalently, $\det (A) -A^{ab}l_a l_b=0$, as one 
finds after some algebra.

\subsection{$A$ and $B$ singular}
Finally, we analyse the case in which both $\det B=0$  \emph{and} $\det{A}=0$.
For simplicity, we will work in the basis in which the symmetric matrix 
$B$ is diagonal. 
Since $B$ is singular, we can choose the basis such that 
$B_0={\rm diag} (0,b_{22},b_{33})$. After inserting this ansatz into 
equations (\ref{eins})-(\ref{drei}), one finds that at least one of the 
two eigenvalues of $B$ must vanish, otherwise there is no real solution. 
Furthermore, at least one of the eigenvalues of $B$ must be different from
zero. 
Otherwise, if $B=0$, eq.\ (\ref{eins}) would imply $C^2=-1_3$ which has no 
solution with a real $3\times 3$ matrix. 
Then, we choose the basis such that $B_0={\rm diag}
(0,0,b_{33})$, 
with $b_{33}\neq 0$. 
We denote the solution in this particular basis as $A_0,B_0$ and $C_0$. 
Using this form of the matrix $B$, one is able to find, after some algebra, 
that the equations (\ref{eins})-(\ref{drei}) admit the following 
{\em four parameter} solution:
\begin{equation}\label{a3}
A_0= \left( 
\begin{array}{ccc}
-c_{23}^2c_{12}^2b_{33}^{-1} & c_{13}c_{23}b_{33}^{-1} &  
-c_{12}c_{23}b_{33}^{-1} \\
c_{13}c_{23}b_{33}^{-1} &  - c_{13}^2 c_{12}^{-2}b_{33}^{-1} 
& c_{13}c_{12}^{-1} b_{33}^{-1} \\
-c_{12}c_{23}b_{33}^{-1} &  c_{13}c_{12}^{-1} b_{33}^{-1} & -b_{33}^{-1} 
\end{array} \right) \, ,
\end{equation}
\begin{equation}\label{b3}
B_0= \left( 
\begin{array}{ccc}
0 & 0 & 0 \\
0 & 0 & 0 \\
0 & 0 & b_{33}
\end{array} \right) \, ,
\end{equation}
\begin{equation}\label{c3}
C_0= \left( 
\begin{array}{ccc}
0 & c_{12} & c_{13} \\
-c_{12}^{-1}  & 0 & c_{23} \\
0 & 0 & 0
\end{array} \right) \,.
\end{equation}
No real solution exists if $c_{12}=0$.
The general solution can thus be found from  this special one by a 
similarity transformation:
\begin{equation}\label{simi}
A=SA_0S^{-1} \qquad B=SB_0S^{-1} \qquad C=SC_0S^{-1}\,.
\end{equation}
Here $S$ is an arbitrary regular matrix which keeps $A$ and $B$ symmetric.
This means that $S$ has to be orthogonal (see for instance ref.\  
\cite{fischer}, p.\ 297).
Therefore, the general solution has 7 independent components, 
due to the 3 additional parameters corresponding to the orthogonal 
transformation. These 7 independent components are thus in agreement 
with the 9 independent components of the regular solution, provided one 
takes into account the two conditions $\det{A}=\det{B}=0$.

The orthogonal group $O(3)$ is given by the direct product of parity 
transformations $P$ and the special orthogonal subgroup $SO(3)$. We just 
have to consider $SO(3)$
transformations as parity transformations leave the solution invariant.
We consider the parametrization of $SO(3)$ based on the generators of
rotations with respect to the Cartesian axes,
\begin{equation}
J_1=\left(
\begin{array}{ccc}
0&0&0\\0&0&-i\\0&i&0
\end{array}
\right)\,, \quad
J_2=\left(
\begin{array}{ccc}
0&0&i\\0&0&0\\-i&0&0
\end{array}
\right)\,\mbox{ and }\,
J_3=\left(
\begin{array}{ccc}
0&-i&0\\i&0&0\\0&0&0
\end{array}
\right)\,.
\end{equation}
Any matrix $S \in SO(3)$ is given by
\begin{equation}
S(\theta^a)=\exp{(i J_a \theta^a)}\,.
\end{equation}
Hence, the general solution for this case is given by 
\begin{eqnarray}
A&=&\exp{(i J_a \theta^a)}\, A_0 \, \exp{(-i J_a \theta^a)}\label{a4}\,,\\
B&=&\exp{(i J_a \theta^a)}\, B_0 \, \exp{(-i J_a \theta^a)}\label{b4}\,,\\
C&=&\exp{(i J_a \theta^a)}\, C_0 \, \exp{(-i J_a \theta^a)}\,.\label{c4}
\end{eqnarray}

\section{The extraction of the metric}
\subsection{The regular solution of the closure relation}
For a regular matrix $B$, the metric has been extracted in two independent
ways up to a conformal factor.
The first method \cite{metric1, metric} makes use of two formulas by
Urbantke, see refs.\ \cite{Urban1,Urban2}. 
On the other hand, Obukhov et al.\ \cite{wavemetric} have studied 
the propagation of electromagnetic waves for the most general case of a 
linear 
constitutive relation, see (\ref{dual}). They found that the wave covector, 
$q_i$, describing the propagation of wave fronts, fulfills a fourth order 
equation
\begin{equation}\label{4metric}
G^{ijkl}\,q_iq_jq_kq_l=0 \, ,
\end{equation}
with
\begin{equation}\label{gijkl}
G^{ijkl}:=\frac{1}{4!}\chi^{mnp(i}\,\chi^{j|qr|k}\,\chi^{l)stu}\,
\epsilon_{mnrs}\,\epsilon_{pqtu} \, .
\end{equation}
This fourth order Fresnel equation is found to reduce to an equation of 
second order in the case of the regular solution (\ref{s1}) of the closure 
relation (\ref{cr1}), i.e.
\begin{equation}\label{lightcone}
G^{ijkl}\,q_iq_jq_kq_l=\sqrt{|g|}\,(g^{ij}q_iq_j)^2\, .
\end{equation}
The factor $\sqrt{|g|}$, with $g:=\det (g_{ij})$, is necessary since 
$\chi^{ijkl}$ is a tensor density of weight $+1$ so that $G^{ijkl}$ is 
also a tensor density of weight $+1$.

As $g^{ij}\,q_iq_j=0$ defines the lightcone at each event, one can read off 
the metric coefficients from (\ref{lightcone}), up to a conformal factor.
In this case, the resulting metric, obtained by the two different methods,
reads
\begin{equation}
 g^{ij}= \frac{1}{\sqrt{\det B}}\left(\begin{array}{c|c}  
  1- (\det B)^{-1} k_c k^c & -k^b\\ \hline  
  -k^a & -(\det B)(B^{-1})^{ab}\end{array} \right)\, , 
\end{equation} 
with $k_a:=B_{ab}\,k^b$.  It can be shown that $g^{ij}$ has Lorentzian 
signature. For details see refs.\ \cite{metric1,metric}.

\subsection{Solution for regular $A$ and singular $B$}
After evaluating (\ref{4metric}) and (\ref{gijkl}) for the solution 
(\ref{s2}), we find that the Fresnel equation (\ref{4metric}) also separates. 
From it, 
we read off the components of the metric up to a conformal factor. We find
\begin{equation}
g^{ij}=\left(\begin{array}{c|c}  
\det A & l^b\\ \hline  
 l^a & (\det A)^{-1}\,l^al^b- A^{ab}\end{array}
\right)\, ,
\end{equation} 
where we have defined $l^c:=A^{ac}l_a$. Again, this metric has Lorentzian 
signature, since $g:=\det{(g_{ij})}=-\det{A}^{-2}<0$.

\subsection{The degenerated case}
\noindent
Finally, for our special solution (\ref{a3})-(\ref{c3}), we find that 
(\ref{4metric}) also separates. The corresponding metric, up to a conformal 
factor, is found to be 
\begin{equation}
g_0^{ij}=\left(\begin{array}{cccc}
0 & c_{23}c_{12}^2 & -c_{13} & c_{12}\\
c_{23}c_{12}^2 & b_{33}c_{12}^2 & 0 & 0 \\
-c_{13} & 0 & b_{33} & 0\\
c_{12}& 0 & 0 & 0
\end{array}\right)\,.\label{gspec}
\end{equation}
The determinant in this case is $g=-b_{33}^{-2}c_{12}^{-4}<0$, so that the 
metric has Lorentzian signature, as in the previous cases.

In order to study the case of solution (\ref{simi}), we look how the 
quantities, in particular $G^{ijkl}$ as defined in (\ref{4metric}), change 
under the orthogonal transformation\footnote{Alternatively, 
one can consider transformations of the coframe basis, as pointed out by 
Obukhov \cite{book,obuk}.} $S$.  
  
From (\ref{block}) and (\ref{simi}) one can write the corresponding 
transformed spacetime tensor density as 
\begin{equation}
\chi^{ijkl}=\Lambda^i_{\ p}\Lambda^j_{\ q}\Lambda^k_{\ r}\Lambda^l_{\ s}\, 
\chi^{pqrs}_0 \, ,
\end{equation}
where we have defined 
\begin{equation}
\Lambda^i_{\ j}:=\left(
  \begin{array}{c|c}
  1 & 0 \\
\hline
  0 & S
  \end{array}\right) \, . \label{simi4}
\end{equation}
From (\ref{gijkl}) we find the corresponding expression for the transformed 
tensor density $G^{ijkl}$. Since $\det (\Lambda^i_{\ j})=1$, it takes the simple 
form 
\begin{equation}
  \label{transG}
  G^{ijkl}=\Lambda^i_{\ p}\Lambda^j_{\ q}\Lambda^k_{\ r}\Lambda^l_{\ s}\, 
G^{pqrs}_0 \, .
\end{equation}
Using this transformation law, one can easily prove the following: For
orthogonal transformations (\ref{simi}) and  (\ref{simi4}), the tensor density 
$G^{ijkl}$ reduces the forth order Fresnel equation to the light cone 
equation, see (\ref{lightcone}), 
provided the tensor density $G_0^{ijkl}$ does. Thus, we find 
that the 
resulting transformed metric can be written in terms of the initial one in 
the form:
\begin{equation}
g^{ij}=\Lambda^i_{\ k}\Lambda^j_{\ l}\,g^{kl}_0 \, .
\end{equation}

We do not give here the explicit expression for this metric. We observe however 
that it has 7 independent components, and a Lorentzian signature.

\section{Conclusions}
By completing a recent analysis, we found a $1$-to-$1$ correspondence between
spacetime tensors fulfilling the closure relation and the conformally
invariant part of metric tensors with Lorentzian
signature. 
A new, more instructive proof for the regular solution (\ref{s1}) is given.

In all cases, we checked the following important property:
We took the conventional Hodge dual of a $2$-form with respect to the metric 
we derived.
This coincides with the action of the duality operator $^\#$, defined in
terms of the spacetime relation, fulfilling the closure relation. In other
words, we have derived (\ref{chig}) from our linear ansatz (\ref{linear}),
the latter of which is constrained by the closure relation (\ref{cr1}). 

This result, previously found only for
regular sub-matrices $A$ and $B$, is then valid for all solutions. 

\subsection*{Acknowledgments}
We are grateful to Friedrich W. Hehl for useful comments and discussion of
the results, and to Yuri N. Obukhov for some critical and helpful
remarks. AG thanks C. Heinicke for some help concerning computer algebra. GFR
would like to thank the German Academic Exchange Service (DAAD) for a
graduate fellowship (Kennziffer A/98/00829).
   

\begin{thebibliography}{}
\bibitem{metric1}
Y.\ N.~Obukhov, F.\ W.~Hehl, 
{\it Space-time metric from linear electrodynamics}, 
{\sl Phys.\ Lett.} {\bf B 458} (1999) 466-470.
Los Alamos Eprint Archive: gr-qc/9904067. 

\bibitem{book} F.\ W.\ Hehl, Y.\ N.\ Obukhov, {\em Foundations of classical
electrodynamics}. Birkh\"auser, Boston, MA (2001/02).

\bibitem{foundem}  F.\ W.\ Hehl, Y.\ N.\ Obukhov, G.\ F.\ Rubilar, {\it 
Classical Electrodynamics: A Tutorial on its
Foundations}, in {\sl Quo vadis geodesia...?, Festschrift f\"ur E.
Grafarend}. F.\ Krumm, V.\ S.\ Schwarze (Eds.), Stuttgart (1999).
Los Alamos Eprint Archive: physics/9907046.

\bibitem{post} E.\ J.\ Post, {\em Formal Structure of Electromagnetics}
-- General Covariance and Electromagnetics. North Holland, 
Amsterdam (1962) and Dover, Mineola, New York (1997).

\bibitem{gordon} W.\ Gordon, {\itshape Zur Lichtfortpflanzung nach der
Relativit\"atstheorie}, {\sl Annalen der Physik} {\bf 72} (1923) 421-456.

\bibitem{leonhardt} U.\ Leonhardt, P.\ Piwnicki, {\itshape Relativistic 
effects of
light in moving media with extremely low group velocity}, {\sl Phys.\ Rev.\
Lett.} {\bf 84} (2000) 822; {\bf 85} (2000) 5253; 
M.\ Visser, {\sl Phys.\ Rev.\ Lett.} {\bf 85} (2000) 5252.

\bibitem{lorenci} V.\ A.\ De Lorenci, M.\ A.\ Souza, {\itshape Electromagnetic
wave propagation inside a material medium: an effective geometry
interpretation},
Los Alamos Eprint Archive: gr-qc/0102022 (2001).

\bibitem{lorenci2}
V.\ A.\ De Lorenci, R.\ Klippert, M.\ Novello, J.\ M.\ Salim, {\itshape Light 
propagation in non-linear electrodynamics}, 
{\sl Phys.\ Lett.} {\bf B 482} (2000) 134-140.
Los Alamos Eprint Archive: gr-qc/0005049.

\bibitem{metric} F.\ W.\ Hehl, Y.\ N.\ Obukhov, G.\ F.\ Rubilar, {\it 
Spacetime metric from linear eletrodynamics II}, {\sl Annalen der Physik} 
{\bf 11} (2000) Spec.\ Iss.\ SI 71 - SI 78.
Los Alamos Eprint Archive: gr-qc/9911096.

\bibitem{energy-momentum} F.\ W.\ Hehl, Y.\ N.\ Obukhov, {\itshape On the
energy-momentum current of the electromagnetic field in a pre-metric
axiomatic approach}, Los Alamos Eprint Archive: gr-qc/0103020.

\bibitem{mielke} E.\ Mielke, {\itshape Geometrodynamics of gauge fields},
Akademie-Verl., Berlin (1981).

\bibitem{fischer} G.\ Fischer, {\itshape Lineare Algebra},
Vieweg, Braunschweig (1997) (11$^{\rm th}$
Ed.).

\bibitem{Urban1} H.\ Urbantke, {\em A quasi-metric associated with
$SU(2)$ Yang-Mills field}, {\sl Acta Phys. Austriaca Suppl.} {\bf 29}
(1978) 875.
 
\bibitem{Urban2} H.\ Urbantke, {\em On integrability properties of
$SU(2)$ Yang-Mills fields. I. Infinitesimal part}, {\sl J.\ Math.\
Phys.} {\bf 25} (1984) 2321-2324.

\bibitem{wavemetric} Y. N. Obukhov, T. Fukui, G. F. Rubilar, {\it Wave
propagation in linear electrodynamics}, {\sl Phys. Rev.} {\bf D 62} (2000) 
044050.
Los Alamos Eprint Archive: gr-qc/0005018.

\bibitem{obuk}Y. N. Obukhov, private communication (2001).
\end{thebibliography}
\end{document}